\documentclass[prd,aps,nofootinbib,preprintnumbers]{revtex4}
\usepackage[T2A]{fontenc}
\usepackage{amsmath,amssymb}
\usepackage{eucal}
\usepackage{color}
\usepackage[dviwindo]{graphicx}
\newcommand{\bs}{\begin{subequations}}
\newcommand{\es}{\end{subequations}}
\numberwithin{equation}{section}
\newcommand{\ben}{\begin{eqnarray}}
\newcommand{\een}{\end{eqnarray}}
\newcommand{\la}{\label}

\begin{document}

\title{Riemannian $\mathbf{(1+d)}$-Dim Space-Time Manifolds with Nonstandard Topology\\
which Admit Dimensional Reduction to Any Lower Dimension \\
and Transformation of the Klein-Gordon Equation \\
to the $\mathbf{1}$-Dim Schr\"odinger Like Equation}

\author{P. P.  Fiziev
\footnote{fizev@phys.uni-sofia.bg\\
fizev@theor.jinr.ru}}
\affiliation{Department of Theoretical Physics, University of Sofia, Boulevard
5 James Bourchier, Sofia 1164, Bulgaria\\
and\\
BLTF, JINR, Dubna, 141980 Moscow Region, Rusia}

 \begin{abstract}
This rather technical paper presents some generalization of the results
of recent  publications \cite{Shirkov2010, DVPF2010, PFDV2010} where toy models of dimensional reduction
of space-time were considered.

Here we introduce and consider a specific type of multidimensional space-times
with nontrivial topology and nontrivial Riemannian metric, which admit a reduction
of the dimension $d$ of the space to any lower one $d_{low} = 1, 2, \dots, d-1$.
The variable geometry is described by several variable radii of compactification
of part of space dimensions.

We succeed once more in transforming the shape of the variable geometry of the $d$-dimensional spaces under consideration to a specific
potential interaction, described by the potential $V$ in the one-dimensional Schr\"odinger-like equation.

This way one may hope to study the possible physical signals going from both higher and lower dimensions
into our obviously four dimensional real world.
\end{abstract}

\sloppy
\maketitle
\section{Introduction}
\begin{figure}[htbp]
\centering
\begin{minipage}{11.cm}
\vskip -3.2truecm
\hskip -4.5truecm
\includegraphics[width=3.truecm, viewport=4 1 100 220]{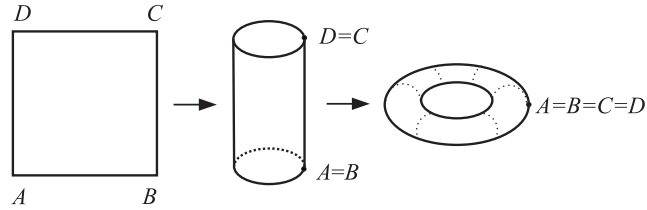} \vskip .5truecm
\caption{\small Obtaining torus from a square by gluing the corresponding boundaries
\hskip .5truecm -- a graphical representation
of the double-periodic boundary conditions}
\label{Fig1}
\end{minipage}
\end{figure}

First, let us consider a $(1+3)$-Dim manifold $\mathbb{M}^{(1,3)}_{t \phi_1 \phi_2 z}$ as a hypersurface in
a flat pseudo-Euclidean $(1+5)$-Dim space $\mathbb{E}^{(1,5)}_{x^0 x^1 x^2 x^3 x^4 x^5}$ with signature $\{+,-,-,-,-,-,\}$
defined by the equations:
\ben
\mathbb{M}^{(1,3)}_{t \phi_1 \phi_2 z}:
\left\{\begin{array}{cccc}
x^0=t,\,\,\,\,x^1=\rho_1(z)\cos\phi_1, \,\,\,\,x^3=\rho_2(z)\cos\phi_2,\cr
x^5=z,\,\,\,\,x^2=\rho_1(z)\sin\phi_1, \,\,\,\,x^4=\rho_2(z)\cos\phi_2,
\la{M}
\end{array}
\right.
\een
assuming $t\in(-\infty,\infty)$, $z\in(-\infty,\infty)$, and $\phi_{1,2}\in [0,2\pi]$.
It is obvious from \eqref{M} that the manifold $\mathbb{M}^{(1,3)}_{t \phi_1 \phi_2 z}$
has a structure $\mathbb{M}^{(1,3)}_{t \phi_1 \phi_2 z}=\mathbb{R}^{(1)}_t\otimes\mathbb{T}^{(2)}_{\phi_1\phi_2}\otimes\mathbb{R}^{(1)}_z$,
$\mathbb{T}^{(2)}_{\phi_1\phi_2}$ being the torus $\mathbb{T}^{(2)}_{\phi_1\phi_2}=\mathbb{S}^{(1)}_{\phi_1}\otimes\mathbb{S}^{(1)}_{\phi_2}$,
see Fig.\ref{Fig1}.

Physically this means that we consider the square $\{\phi_1,\phi_2\}\in [0,2\pi]\otimes[0,2\pi]$
as a flat domain with periodic boundary conditions for any field
$\Psi(t,\phi_1,\phi_2,z) \equiv \Psi(t,\phi_1+2 n_1\pi,\phi_2+2 n_2\pi,z)$,
$n_1,n_2$ being arbitrary integers. Thus we compactify some of the space dimensions, but instead of fixing
the radii of compactification $\rho_1(z)$ and $\rho_2(z)$,
we let them to depend on the non-compactified coordinate $z$.
In the domains of very big values of the radii $\rho_1(z)$ and $\rho_2(z)$ the space will look like the usual flat $3$-Dim space.
If one or two of the radii of compactification $\rho_1(z)$ and $\rho_2(z)$ become very small, the dimension
of the $3$-Dim space effectively reduces to a lower one.

The manifold \eqref{M} is an obvious generalization of the $(1+2)$-Dim manifolds with cylindrical symmetry, considered in
\cite{Shirkov2010, DVPF2010, PFDV2010}. There we had $x^4$ and $x^5$ with fixed values, say zero.

\section{The Induced Riemannian Geometry of the Manifold $\mathbb{M}^{(1,3)}_{t \phi_1 \phi_2 z}$ }

The restriction of the $6$-Dim (pseudo)Euclidean interval ${}^{(6)}\!ds^2=(dx^0)^2-(dx^1)^2-(dx^2)^2-(dx^3)^2-(dx^4)^2-(dx^5)^2$
on the manifold \eqref{M} induces the following simple (pseudo)Riemannian $4$-Dim interval
\ben
ds^2=dt^2-\rho_1 (z)^2 d \phi_1^2 -\rho_2(z)^2 d\phi_2^2 -\left(1+\rho_1^\prime(z)^2+\rho_2^\prime(z)^2\right)dz^2.
\la{ds2}
\een

The $3$-Dim (to be physical) space with Riemannian interval
\ben
dl^2=\rho_1 (z)^2 d \phi_1^2 +\rho_2(z)^2 d\phi_2^2 +\left(1+\rho_1^\prime(z)^2+\rho_2^\prime(z)^2\right)dz^2
\la{dl2}
\een
has a nontrivial scalar curvature
\ben
{}^{(3)}\!R=&2&{\frac{\big(\rho_2(z)\left((1+\rho_2^\prime(z)^2\right))-\rho_1(z)\rho_1^\prime(z)\rho_2^\prime(z)\big)\rho_1^{\prime\prime}(z)}
{\rho_1(z)\rho_2(z)\big(1+(\rho_1^\prime(z))^2+(\rho_2^\prime(z))^2\big)^2}}+\nonumber\\
+&2&{\frac{\big(\rho_1(z)\left((1+\rho_1^\prime(z)^2\right))-\rho_2(z)\rho_1^\prime(z)\rho_2^\prime(z)\big)\rho_2^{\prime\prime}(z)}
{\rho_1(z)\rho_2(z)\big(1+(\rho_1^\prime(z))^2+(\rho_2^\prime(z))^2\big)^2}}+\nonumber\\
+&2&{\frac {\rho_1^\prime(z)\rho_2^\prime(z)}
{\rho_1(z)\rho_2(z)\big(1+(\rho_1^\prime(z))^2+(\rho_2^\prime(z))^2\big)}},
\la{3R}
\een
but its $3$-Dim Weyl tensor vanishes identically. Since we deal with a nontrivial $3$-Dim Riemannian  manifold,
the vanishing of the Weyl tensor is not sufficient to conclude that the physical space with metric \eqref{dl2} is conformally flat.
The necessary and sufficient condition for conformal flatness of $3$-Dim space is to vanish its Taub tensor \cite{Taub}
$${}^{(3)}T_{\alpha\beta\gamma}=\triangledown_\alpha\left({}^{(3)}\!R_{\beta\gamma}-(1/4){}^{(3)}\!R\,{}^{(3)}\!g_{\beta\gamma}\right)-
\triangledown_\beta\left({}^{(3)}\!R_{\alpha\gamma}-(1/4){}^{(3)}\!R\,{}^{(3)}\!g_{\alpha\gamma}\right).$$
In general, for the metric in \eqref{dl2} the Taub tensor does not vanish.
It is easy to check that the $4$-Dim pseudo-Riemannian manifold with metric \eqref{ds2} is not conformally flat, too,
since its  $4$-Dim Weyl tensor does not vanish.

In the space-times at hand the $4$-Dim scalar curvature coincides with the $3$-Dim one \eqref{3R}, i.e., ${}^{(4)}\!R={}^{(3)}\!R$.

The obtained $3$-Dim metric is diagonal with diagonal elements $\rho_1(z)^2$, $\rho_2(z)^2$, and $\rho_3(z)^2=1+(\rho_1^\prime(z))^2+(\rho_2^\prime(z))^2$.
The square root of its determinant is $\sqrt{{}^{(3)}\!g(z)}=\rho_1(z)\rho_2(z)\rho_3(z)=\sqrt{{}^{(4)}\!g(z)}$.

The natural (orthogonal) tetrad basis $e^{(a)}_\alpha(z)$ $\big(g_{\alpha\beta}(z)=e^{(a)}_\alpha(z) e^{(b)}_\beta(z) \eta_{ab}\big)$
for the $4$-Dim interval \eqref{ds2} is needed to construct the Dirac equation,
see for example the recent articles \cite{Teryaev2005,Teryaev2009,Neznamov} and the references therein.
In our case this basis is a very simple one:
\ben
e^{(0)}_\alpha(z)=
\left(\begin{array}{c}
1\\0\\0\\0
\end{array}\right),\,\,\,\,
e^{(1)}_\alpha(z)=
\left(\begin{array}{c}
0\\ \rho_1(z)\\0\\0
\end{array}\right),\,\,\,\,
e^{(2)}_\alpha(z)=
\left(\begin{array}{c}
0\\0\\ \rho_2(z)\\0
\end{array}\right),\,\,\,\,
e^{(3)}_\alpha(z)=
\left(\begin{array}{c}
0\\0\\0\\\rho_3(z)
\end{array}\right).
\la{tetrads}
\een
Since the $3$-Dim space is not conformally flat, the methods for studying the Dirac equation used in \cite{Teryaev2005,Teryaev2009}
are not directly applicable to space-times with metric \eqref{ds2}
if one does not consider the very special case $\rho_1(z)=\rho_2(z)=\rho_3(z)$.

Obviously, taking the limit $\rho_1(z)\to 0$, or the limit $\rho_2(z)\to 0$, we are able to make a reduction of dimension of the
physical space from $d=3$ to $d=2$. The simultaneous limits $\rho_1(z)\to 0$ and $\rho_2(z)\to 0$ will bring us
to a one-dimensional physical space ($d=1$).
Hence, working with the toy-metric \eqref{ds2}, we are able to study different physical phenomena,
related with dimensional reduction \cite{Shirkov2010, DVPF2010, PFDV2010}
exploring all physically interesting lower dimensions: $d_{low}=1,2$.

\section{The $3$-Dim Laplacian, new variables and reduction of the Klein-Gordon equation to the $\mathbf{1}$-Dim Schr\"odinger like one}
In the coordinates $\phi_1,\phi_2,z$ the $3$-Dim Laplacian reads
\ben
\Delta_3={\frac 1 {\rho_1(z)^2}}\partial^2_{\phi_1}+{\frac 1 {\rho_2(z)^2}}\partial^2_{\phi_2}+
{\frac 1 {\rho_1(z)^2\rho_2(z)^2}}\left({\frac{\rho_1(z)\rho_2(z)}{\rho_3(z)}}\partial_z\left({\frac{\rho_1(z)\rho_2(z)}{\rho_3(z)}}\partial_z\right)\right).
\la{3Lapalace_z}
\een
The introduction of the new variable
\ben
u=u(z)=\int{\frac{\rho_3(z)}{\rho_1(z)\rho_2(z)}}dz=\int{\frac{\sqrt{1+\rho_1^\prime(z)^2+\rho_2^\prime(z)^2}}{\rho_1(z)\rho_2(z)}}dz
\la{u}
\een
is analogous to the one used in \cite{PFDV2010}\footnote{We obtain the previous result \cite{PFDV2010} putting $\rho_1(z)=\rho(z),\,\phi_1=\phi$
and fixing the values of the $\rho_2(z)=1$ and of the angle $\phi_2$, say $\phi_2=0$.} and simplifies the form of the $3$-D Laplacian:
\ben
\Delta_3={\frac 1 {\varrho_1(u)^2}}\partial^2_{\phi_1}+{\frac 1 {\varrho_2(u)^2}}\partial^2_{\phi_2}+
{\frac 1 {\varrho_1(u)^2\varrho_2(u)^2}}\partial_u^2, \quad\text{where}\quad  \varrho_{1,2}(u)=\rho_{1,2}\big(z(u)\big).
\la{3Lapalace_u}
\een

Consider the standard Klein-Gordon equation
\ben
\la{KGE}
 \left(\Box-M^2\right)\varphi=0, \quad\text{where}\quad \Box=-\partial^2_{tt} +\Delta_3.
 \een

It admits a separation of the variables $\varphi(t,\phi_1,\phi_2,z)=T(t)
 \Phi_1(\phi_1) \Phi_2(\phi_2) Z(z)$, yielding a system of ordinary differential
 equations (ODEs). Three of them are simple:
$T^{\prime\prime}+\omega^2 T=0,\,\,\,\Rightarrow\,\,\, T(t)=
 e^{-i\omega t}$ and
 $\Phi_1^{\prime\prime}+m_1^2\Phi_1=0,\quad\Rightarrow\,\,\,
 \Phi_1(\phi_1)=e^{im_1\phi_1},\,\,\,m_1=0, \pm1,\pm2,\dots$,
$\Phi_2^{\prime\prime}+m_2^2\Phi_2=0,\quad\Rightarrow\,\,\,
 \Phi_2(\phi_2)=e^{im_2\phi_2},\,\,\,m_2=0, \pm1,\pm2,\dots$
The only nontrivial equation is the one for the function $Z(z)$.
Its explicit form
\ben\label{Z_z}
{\frac 1 {\rho_1(z)^2\rho_2(z)^2}}\left({\frac{\rho_1(z)\rho_2(z)}{\rho_3(z)}}\partial_z\left({\frac{\rho_1(z)\rho_2(z)}{\rho_3(z)}}\partial_z Z\right)\right)+
 \left(\!\omega^2-M^2\!-\!{\tfrac {m_1^2} {\rho_1(z)^2}}\!-\!{\tfrac {m_2^2} {\rho_2(z)^2}}\!\right)Z\!=\!0,
 \\ \Rightarrow\,\, Z(z)=Z(z;\omega,m_1,m_2) \nonumber\la{ode:c}
 \een
recovers the physical meaning of the terms $m_1^2/\rho_1(z)^2$ and $m_2^2/\rho_2(z)^2$.
These describe the potential energy of the centrifugal-like
forces which act for $m_{1,2}\neq 0$. One has to stress that these
terms present a more complicated example of inertial
forces. Such forces are an unavoidable feature of the motion in curved
space-times. The inertial forces will certainly arise in the junction
domains of transition between the parts of space with
different dimensions \cite{PFDV2010}.

After the transition to the variable $u=u(z)$ (see Eq. \eqref{u})
Eq. \eqref{Z_z} acquires a Schr\"odinger like form
\ben
\psi^{\prime\prime}(u)+\big(E-V(u)\big)\psi(u)=0,
\la{Schr}
\een
namely:
\ben\la{eqU}
\psi^{\prime\prime}(u)+\big(-\left(M^2-\omega^2\right)\varrho_1(u)^2\varrho_2(u)^2-m_1^2\varrho_2(u)^2-m_2^2\varrho_1(u)^2\big)\psi(u)=0,
\een
with identification
$E=0,\,V(u)=\left(M^2-\omega^2\right)\varrho_1(u)^2\varrho_2(u)^2+m_1^2\varrho_2(u)^2+m_2^2\varrho_1(u)^2$,
$Z(z)=\psi(u(z))$\footnote{In variable $u$ the $3$-Dim interval \eqref{dl2} acquires the form $dl^2=\rho_1 (z)^2 d \phi_1^2 +\rho_2(z)^2 d\phi_2^2 +
\varrho_1(u)^2\varrho_2(u)^2du^2$.
The tetrad basis \eqref{tetrads} has the same form with $\rho_{1,2}(z)$ replaced with $\varrho_{1,2}(u)$,
and $\rho_3(z)$ replaced with $\varrho_3(u)=\varrho_1(u)\varrho_2(u)$.}.
The relation with the analogous result in \cite{PFDV2010} is described once more in the footnote 1 on page 2.
Hence, we can construct a large class of exactly solvable models, based on Eq. \eqref{eqU}, see \cite{DVPF2010, PFDV2010}.

There exists one more possibility: To consider the conformally invariant Klein-Gordon equation (CIKGE),
discovered  by  Penrose and Chernikov-Tagirov \cite{Penrose, CT}.
The CIGGE in $(1+3)$-Dim reads
\ben
\la{CIKGE}
\left( \Box- {}^{(4)}\!R/6 \right)\varphi=0.
\een

In this case $M=0$ and the potential $V(u)=V_c(u)$ in Eq. \eqref{Schr} has the following more complicated form:
$$V_{c}(u)=\left({}^{(4)}\!R/6-\omega^2\right)\varrho_1(u)^2\varrho_2(u)^2+m_1^2\varrho_2(u)^2+m_2^2\varrho_1(u)^2.$$
Note that in the variable $u$ instead of expression \eqref{3R} we obtain a much simplear one:
\ben
{}^{(3)}\!R={}^{(4)}\!R={\frac 2{\varrho_1^2\varrho_2^2}}
\left(\left({\frac{\varrho_1^\prime}{\varrho_1}}\right)^\prime+
      \left({\frac{\varrho_1^\prime}{\varrho_1}}\right)^\prime
      -{\frac{\varrho_1^\prime}{\varrho_1}}{\frac{\varrho_1^\prime}{\varrho_1}}\right).
\la{3Ru}
\een

\section{A Natural Multidimensional Generalization}

It is easy to obtain a natural generalization of the above results for higher dimensions $d>3$. Indeed, let us consider
a $(1+d)$-Dim manifold $\mathbb{M}^{(1,\, d)}_{t \phi_1\dots \phi_{d-1} z}$ as a hypersurface in
a flat pseudo-Euclidean $(1+(2d-1))$-Dim space $\mathbb{E}^{(1,2d-1)}_{x^0 x^1\dots x^{2d-1}}$ with signature $\{+,-,-,-,-,-,\}$
defined by the equations:
\ben
\mathbb{M}^{(1,d)}_{t \phi_1 \phi_2 z}:
\left\{\begin{array}{cccc}
x^0&=t,\,\,\,\,x^1&=\rho_1(z)\cos\phi_1, \,\,\dots,\,\,x^{2d-3}=\rho_{d-1}(z)\cos\phi_{d-1},\cr
x^{2d-1}&=z,\,\,\,\,x^2&=\rho_1(z)\sin\phi_1, \,\,\dots,\,\,x^{2d-2}=\rho_{d-1}(z)\cos\phi_{d-1},
\la{M1d}
\end{array}
\right.
\een
assuming $t\in(-\infty,\infty)$, $z\in(-\infty,\infty)$, and $\phi_{1,\dots,d-1}\in [0,2\pi]$.
It is obvious from \eqref{M1d} that the manifold $\mathbb{M}^{(1,\, d)}_{t \phi_1\dots \phi_{d-1} z}$
has a structure $\mathbb{M}^{(1,\, d)}_{t \phi_1\dots \phi_{d-1} z}=\mathbb{R}^{(1)}_t\otimes\mathbb{T}^{(d-1)}_{\phi_1\dots\phi_{d-1}}\otimes\mathbb{R}^{(1)}_z$,
$\mathbb{T}^{(d-1)}_{\phi_1\dots\phi_{d-1}}$ being the torus
$\mathbb{T}^{(d-1)}_{\phi_1\dots\phi_{d-1}}=\mathbb{S}^{(1)}_{\phi_1}\otimes\dots\otimes\mathbb{S}^{(1)}_{\phi_{d-1}}$.
The geometry of this $(d-1)$-Dim torus reflects the multiply-periodic boundary conditions
 on the fields $\Psi$ in the problem at hand:
$\Psi(t,\phi_1,\dots,\phi_{d-1},z) \equiv \Psi(t,\phi_1+2 n_1\pi,\dots,\phi_{d-a}+2 n_{d-1}\pi,z)$,
$n_1,\dots,n_{d-1}$ being arbitrary integers.

The restriction of the $2d$-Dim (pseudo)Euclidean interval ${}^{(2d)}\!ds^2=(dx^0)^2-(dx^1)^2-\dots-(dx^{2d-1})^2$
on the manifold \eqref{M1d} induces the following simple (pseudo)Riemannian $(1+d)$-Dim interval:
\ben
ds^2=dt^2-\rho_1 (z)^2 d \phi_1^2-\dots -\rho_{d-1}(z)^2 d\phi_{d-1}^2 -\rho_{d}(z)^2dz^2,
\la{dds2}
\een
where $\rho_{d}(z)=\sqrt{1+\rho_1^\prime(z)^2+\dots+\rho_{d-1}^\prime(z)^2}$.
For $d>3$ the $d$-Dim  space with Riemannian interval
\ben
dl^2=\rho_1 (z)^2 d \phi_1^2 +\dots +\rho_{d-1}(z)^2 d\phi_{d-1}^2 +\rho_{d}(z)^2dz^2
\la{ddl2}
\een
has nonvanishing Weyl tensor and a quite complicated nonzero scalar curvature.

In the coordinates $\phi_1,\dots\phi_{d-1},z$ the $d$-Dim Laplacian reads
\ben
\Delta_d={\frac 1 {\rho_1(z)^2}}\partial^2_{\phi_1}+\dots+{\frac 1 {\rho_{d-1}(z)^2}}\partial^2_{\phi_{d-1}}+
{\frac 1 {\rho_1(z)^2\dots\rho_{d-1}(z)^2}}\left({\frac{\rho_1(z)\dots\rho_{d-1}(z)}{\rho_d(z)}}\partial_z\left({\frac{\rho_1(z)\dots\rho_{d-1}(z)}{\rho_d(z)}}\partial_z\right)\right).
\la{dLapalace_z}
\een
The introduction of the new variable
\ben
u=u(z)=\int{\frac{\rho_d(z)}{\rho_1(z)\dots\rho_{d-1}(z)}}dz=\int{\frac{\sqrt{1+\rho_1^\prime(z)^2+\dots+\rho_{d-1}^\prime(z)^2}}{\rho_1(z)\dots\rho_{d-1}(z)}}dz
\la{du}
\een
simplifies the form of the $d$-Dim Laplacian:
\ben
\Delta_d={\frac 1 {\varrho_1(u)^2}}\partial^2_{\phi_1}+{\frac 1 {\varrho_2(u)^2}}\partial^2_{\phi_2}+
{\frac 1 {\varrho_1(u)^2\dots\varrho_{d-1}(u)^2}}\partial_u^2, \quad\text{where}\quad  \varrho_{\alpha=1,\dots,d}(u)=\rho_{\alpha=1,\dots,d}\big(z(u)\big).
\la{dLapalace_u}
\een

After the separation of variables in the corresponding KGE of type \eqref{KGE} we obtain the following nontrivial $Z$-equation:
\ben\label{dZ_z}
{\frac 1 {\rho_1(z)^2\dots\rho_{d-1}(z)^2}}\left({\frac{\rho_1(z)\dots\rho_{d-1}(z)}{\rho_d(z)}}\partial_z\left({\frac{\rho_1(z)\dots\rho_{d-1}(z)}{\rho_d(z)}}\partial_z Z\right)\right)+
 \left(\!\omega^2-M^2\!-\!{\tfrac {m_1^2} {\rho_1(z)^2}}\!-\dots-\!{\tfrac {m_{d-1}^2} {\rho_{d-1}(z)^2}}\!\right)Z\!=\!0.\,\,\,
 \\ \Rightarrow\,\, Z(z)=Z(z;\omega,m_1,\dots,m_{d-1}) \nonumber
 \een

The terms $m_1^2/\rho_1(z)^2, \dots, m_{d-1}^2/\rho_2(z)^2$ describe the potential energy of the centrifugal-like
inertial forces which act for $m_{1,\dots,d-1}\neq 0$.

Using the variable $u$ we obtain instead of Eq. \eqref{dZ_z} Schr\"odinger-like equations \eqref{Schr} with potential
\ben
V(u)=\varrho_1(u)^2\dots\varrho_{d-1}(u)^2\Big(\left(M^2-\omega^2\right)+m_1^2/\varrho_1(u)^2+\dots+m_{d-1}^2/\varrho_{d-1}(u)^2\Big).
\la{dV}
\een

To construct CIKGE \eqref{CIKGE} one has to use the space-time scalar curvature in $u$ variable:
\ben
{}^{(1+d)}\!R={}^{(d)}\!R=2
\left(\sum_{\alpha=1}^{d-1}\left({\frac{\varrho_\alpha^\prime}{\varrho_\alpha}}\right)^\prime
      -\sum_{\alpha,\beta=1; \alpha<\beta}^{d-1}{\frac{\varrho_\alpha^\prime}{\varrho_\alpha}}{\frac{\varrho_\beta^\prime}{\varrho_\beta}}\right)\Bigg/
      {\prod_{\alpha=1}^{d-1}\varrho_\alpha^2}.
\la{dRu}
\een
It defines the potential $V_c(u)$ for the corresponding Scr\"odinger like equation \eqref{Schr}:
\ben
V_{c}(u)=\left({}^{(1+d)}\!R/n_d -\omega^2 +
\sum_{\alpha=1}^{d-1}{\frac{m_\alpha^2}{\varrho_\alpha^2}}\right){\prod_{\alpha=1}^{d-1}\varrho_\alpha^2}, \quad\text{where}\quad n_d=4/(1-1/d).
\la{dVc}
\een

Thus, we succeeded once more in transforming the shape of the variable geometry in the $d$-Dim spaces under consideration to a specific
potential interaction, described by the potential \eqref{dV}, or \eqref{dVc} in the Schr\"odinger-like equations \eqref{Schr}.

Taking a different number of limits $\rho_{\mu}(z)\to 0$ for some set of values of the index
$\mu$ we can describe the dimensional reduction phenomena in the space-times at hand
reducing the dimension $d$ of the space to any lower one $d_{low} = 1, 2, \dots, d-1$.
This way one may hope to study the possible physical signals going from both higher and lower dimensions
into our real world which is obviously a four dimensional one.

\begin{acknowledgments}
This paper was inspired by many fruitful discussions with Dmitry Vasilievich Shirkov during our collaboration,
and greatly influenced by his deep physical intuition and useful advices.
He drew my attention to the problem of dimensional reduction of the physical space-time.
It is a pleasure to thank Irina Yaroslavna Aref'eva, and Oleg Valerianovich Teryaev for useful discussions and
the strong pulses for movement in the right direction of investigations.
The discussions with  Vasily Petrovich Neznamov were very stimulating, too.

The research has been partially supported by
the Bulgarian National Scientific Fund under
contracts DO-1-872, DO-1-895 and DO-02-136 and by the Sofia University Scientific
Fund, contract 185/26.04.2010.
\end{acknowledgments}


\end{document}